# M(BO)$_{k+1}$$^-$ Anions : Novel Superhalogens Based on Boronyl Ligands[*]

Ambrish Kumar Srivastava

*Department of Physics, Deen Dayal Upadhyaya Gorakhpur University, Gorakhpur, 273009, Uttar Pradesh, India*

E-mail: aks.ddugu@gmail.com

**Abstract**

Superhalogens have been a subject of continuous attraction for the last four decades due to their unusual structures and interesting applications. In the quest for new superhalogens based on boronyl (BO) ligands, we investigate $M(BO)_{k+1}^{-}$ anions using high-level ab initio CCSD(T) and B3LYP density functional method for M = Li, Na, K, Be, Mg, Ca, B, and Al. Their structures are linear for M = Li, Na, K, trigonal planar M = Be, Mg, Ca, and tetrahedral for M = B, Al. These anions are energetically and thermodynamically stable against various fragmentations. Their superhalogen property has been established due to their higher vertical detachment energy (VDE) than halogen, being in the range 4.44-7.81 eV. For a given $k$, the stability and VDE of $M(BO)_{k+1}^{-}$ anions decrease with the increase in the size of M, which is due to a decrease in the charge delocalization on BO moieties. There exists a linear correlation between frontier orbitals energy gap and VDE with the coefficient, $R^2 = 0.93888$. It was also noticed that BO ligand, due to lower dimerization energy, can be preferably used as an inorganic analog of CN. We believe that these findings should be interesting for researchers working in superatomic chemistry as well as in boron chemistry.

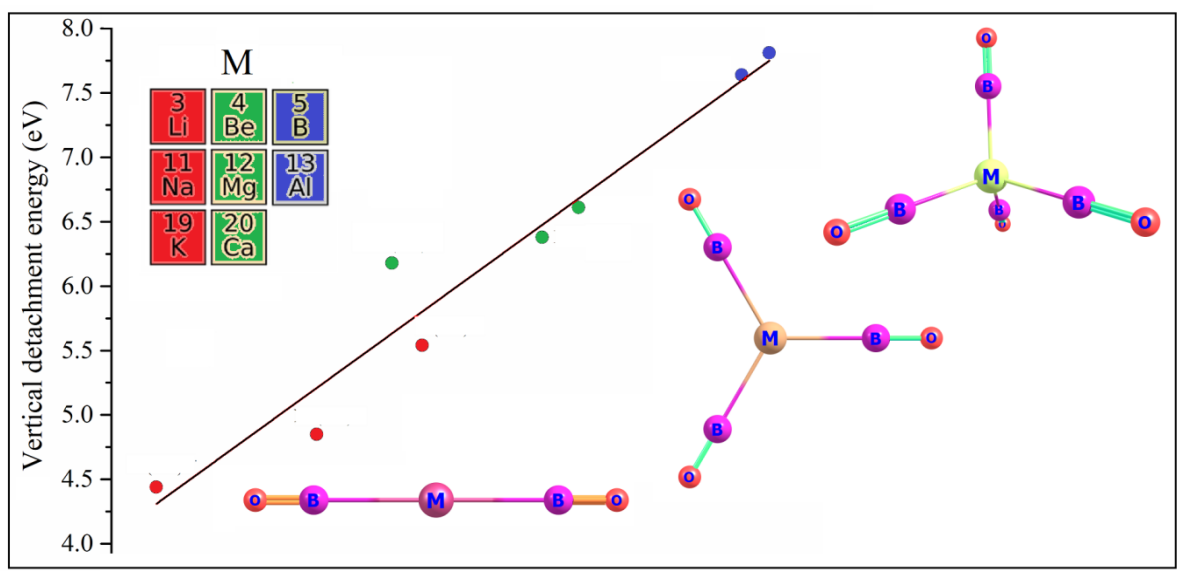

# 1. Introduction

In 1981, the concept of superhalogen was introduced by Gutsev and Boldyrev [1, 2]. In their pioneering work, they designed hypervalent clusters, $ML_{k+1}$, which include a central metal core (M) having valence $k$ with peripheral electronegative ligands (L) such as F, Cl, O, etc and noticed that these clusters possess higher electron affinities (EAs) than halogen, hence the name 'superhalogen'. These unusual species form stable anions whose vertical detachment energy (VDE) is larger than the EA of halogen. The first experimental confirmation of these anions was made by Wang et al. in 1999 [3]. This led to the exploration of new superhalogens using different transition metals as central core [4-7] and various ligands other than F, Cl, or O such as halogenoids [8], acidic groups [9], other ligands [10] or a combination such as OF [11, 12]. The research on superhalogens is not only confined to their unusual structures and properties but also spanned to their several applications. The recent studies reported the application of superhalogens in the design of superacids [13, 14], energy storage materials [15, 16], ionic liquids [17], etc. The ever-increasing scope of superhalogens motivates the research on the design of new superhalogens. In this work, we propose a series of new superhalogen anions, $M(BO)_{k+1}^-$ using boronyl ligands.

Boronyl (BO) has got attention due to its primary role in the oxidation of boron and related systems [18]. $BO^-$ has been widely studied by various spectroscopic means including photoelectron spectroscopy [19] and infrared spectroscopy [20]. There have been extensive studies on small $(BO)_n$ clusters [20-25] both in their neutral and anionic states. BO motif has been employed to design several boron oxide clusters anions such as $B(BO)_2^-$ and $B(BO)_3^-$ [26], $B_2(BO)_2^-$ [27], $B(BO)_4^-$ [28], $B_3O_n^-$ ($n$ = 2-4) [29], etc. Apart from this, several other boronyl based species such as gold boronyl clusters $Au_nBO^-$ ($n$ = 1-3) [30], carbon boronyl clusters $C_n(BO)_n$ ($n$ = 3-7) [31] and compounds $C_2(BO)_mH_n$ [32] have been reported. Despite the enormous progress in the research on boronyls, there is no report on the use of BO as a ligand to induce the superhalogen properties. This motivated us to perform a systematic study on $M(BO)_{k+1}^-$ anions for M = Li, Na, K, Be, Mg, Ca, B, and Al. The objective of this study is to investigate whether the BO can act as a ligand, just as F or CN, to design new superhalogen anions. If yes, whether the potency of BO ligand can compete with F or CN as far as the design of superhalogen is concerned. To answer these questions, we have performed density functional theory and high-level *ab initio* method-based calculations.

## 2. Computational methods

The possible structures of $M(BO)_{k+1}^-$ anions for M = Li, Be, and B were initially optimized by hybrid functional B3LYP [33, 34] with a 6-311+G(d) basis set. The lowest energy structures were re-optimized at B3LYP/6-311+G(3df) level to obtain the equilibrium structures of $M(BO)_{k+1}^-$ anions for M = Li, Na, K, Be, Mg, Ca, B, and Al. The optimization procedure was followed by vibrational frequency calculations to ensure that all optimized structures contain all positive frequency values and thus, correspond to true minima. In order to obtain more accurate energy values, we have performed single-point CCSD(T) [35] calculations over B3LYP/6-311+G(3df) optimized geometries. These CCSD(T) calculated energy values were used to calculate the vertical detachment energy (VDE) of $M(BO)_{k+1}^-$ anions and their dissociation energy values. The partial charges on BO moiety were computed by the natural population analysis (NPA) [36]. All these computations were performed with the aid of the Gaussian 09 [37] program. The validation of the present computational scheme is discussed below.

## 3. Results and discussion

The equilibrium structures of $(BO)_n^-$ clusters are displayed in Fig. 1 and corresponding parameters are listed in Table 1. The bond-length of BO⁻ 1.231 Å and its VDE 2.44 eV agree well with corresponding experimental values of 1.236 Å [38] and 2.51 eV [19] measured by photoelectron spectroscopy. The lowest energy structure of $B_2O_2^-$ is taken from Li et al. [32], whose B-O bond length (1.240 Å) is in accordance with the previously reported value (1.242 Å) at B3LYP/aug-cc-pVTZ level [32] and its VDE is calculated to be merely 0.62 eV. This low VDE may advocate the strong tendency of BO to form a dimer, like F, Cl, CN, SCN, etc. The structures of $B_3O_3^-$ and $B_4O_4^-$ are adopted from previous studies of Zhao et al. [22] and Tian et al. [23]. Their VDEs are also consistent with their observed values [22, 23]. These results suggest that our computational methods are capable of providing reliable data.

Table 1. CCSD(T)//B3LYP/6-311+G(3df) calculation

| Species | BO bond lengths (Å) | Vertical detachment energy, VDE (eV) | |
|---|---|---|---|
| | | Calculated value | Experimental value |
| $BO^-$ | 1.231 | 2.44 | 2.51[a] |
| $B_2O_2^-$ | 1.240 | 0.62 | - |
| $B_3O_3^-$ | 1.215, 1.215, 1.218 | 4.30 | 4.20[b] |
| $B_4O_4^-$ | 1.220, 1.220, 1.226 | 2.85 | 2.81[c] |

[a]Ref. [19]
[b]Ref. [22]
[c]Ref. [23]

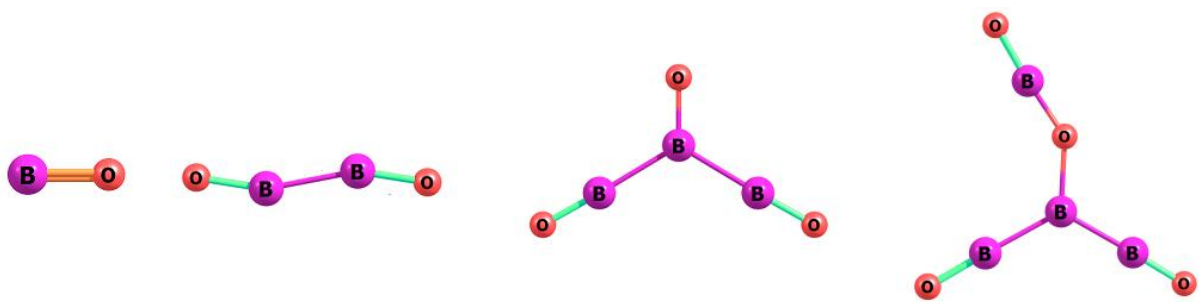

Fig. 1. B3LYP/6-311+G(3df) optimized lowest energy structures of $(BO)_n^-$ anions ($n$ = 1-4).

### 3.1. $M(BO)_{k+1}^-$ anions

In order to study the complexes of BO with M, we have considered various possible initial structures for M = Li, Be, and B. After full optimization, the local minimum structures obtained at the B3LYP/6-311+G(d) level are displayed in Fig. 2. For $Li(BO)_2^-$ and $Be(BO)_3^-$, we obtain bent $C_s$ isomers, which are 1.70 eV and 0.82 eV higher in energy than the corresponding lowest energy structures. Likewise, we obtain a few isomers of $B(BO)_4^-$ lying in the range 2.74-4.59 eV higher than its lowest energy structure (see Fig. 2). A previous study [28] reported some other isomers of B(BO)$_4^-$ with higher-order stationary points (a few imaginary frequencies).

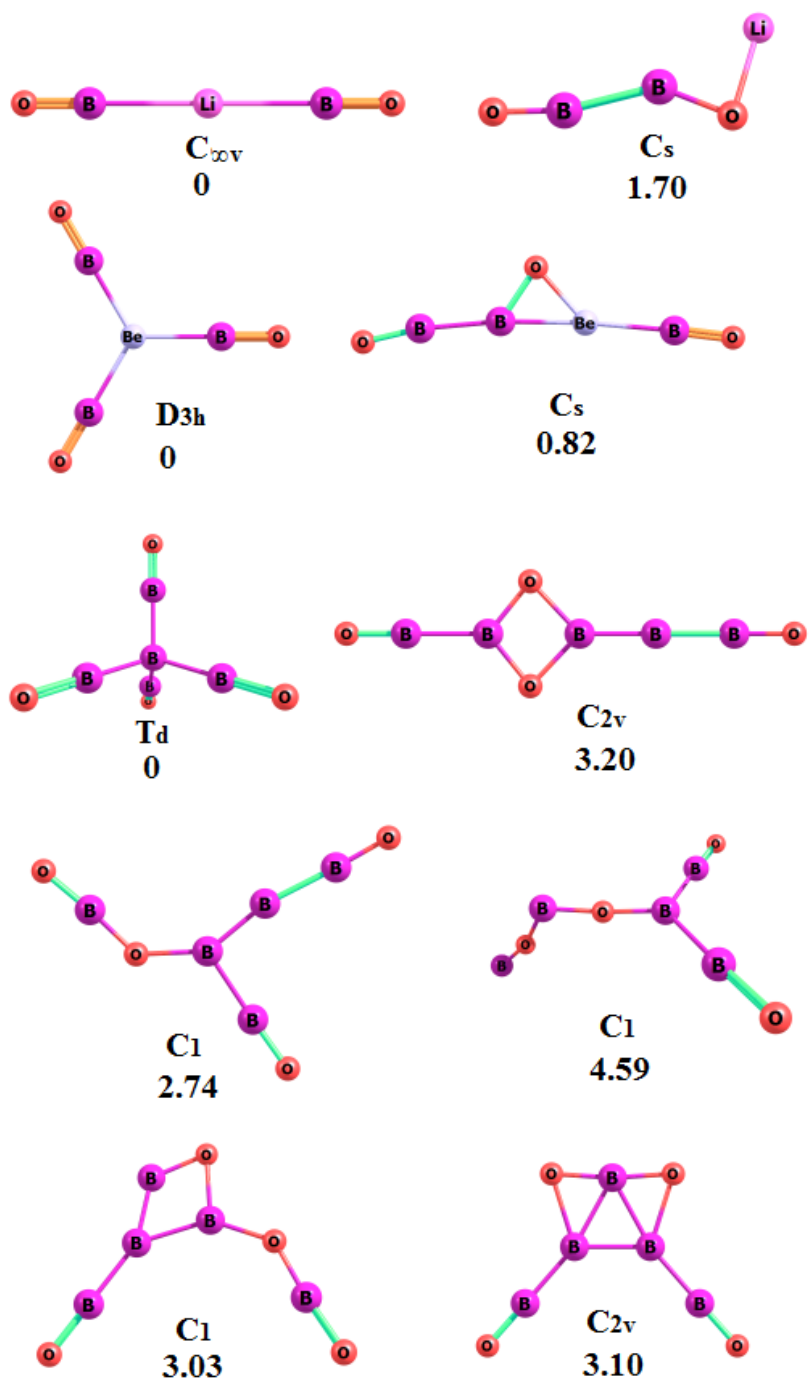

Fig. 2. B3LYP/6-311+G(d) optimized structures of Li(BO)$_2^-$, Be(BO)$_3^-$, and B(BO)$_4^-$ anions. Symmetry and relative energy of isomers (in eV) are also displayed.

Nevertheless, the minimum energies of $M(BO)_{k+1}^-$ correspond to the structure in which all BO moieties interact with central atom M. These structures were re-optimized at B3LYP/6-311+G(3df) level for M = Li, Na, K, Be, Mg, Ca, B, and Al as shown in Fig. 3. Below we discuss these equilibrium structures, their stability, and superhalogen property.

Table 2. The bond length, frontier orbital energy gap ($E_{gap}$), NPA charge on BO ($q$(BO)), Gibbs' free energy change ($\Delta G^{298}$) in $M(BO)_{k+1}^-$ anions. CCSD(T)//B3LYP/6-311+G(3df) calculated vertical detachment energy (VDE) and dissociation energy ($\Delta E$) are listed.

| $M(BO)_{k+1}^-$ anion | Bond length (Å) | | Fragmentation | | | $E_{gap}$ (eV) | VDE (eV) | $q$(BO) (e) |
|---|---|---|---|---|---|---|---|---|
| | B-O | M-B | Path | $\Delta E$ (eV) | $\Delta G^{298}$ (eV) | | | |
| $Li(BO)_2^-$ | 1.221 | 2.230 | $Li + B_2O_2^-$ | 3.09 | 2.84 | 5.65 | 5.54 | -0.65 |
| | | | $Li + BO + BO^-$ | 5.90 | 5.45 | | | |
| $Na(BO)_2^-$ | 1.222 | 2.562 | $Na + B_2O_2^-$ | 2.30 | 2.07 | 4.92 | 4.85 | -0.72 |
| | | | $Na + BO + BO^-$ | 5.10 | 4.68 | | | |
| $K(BO)_2^-$ | 1.224 | 3.012 | $K + B_2O_2^-$ | 1.94 | 1.82 | 3.81 | 4.44 | -0.80 |
| | | | $K + BO + BO^-$ | 4.75 | 4.42 | | | |
| $Be(BO)_3^-$ | 1.215 | 1.885 | $Be + B_3O_3^-$ | 3.22 | 3.44 | 6.73 | 6.61 | -0.27 |
| | | | $Be + BO + B_2O_2^-$ | 7.75 | 7.19 | | | |
| | | | $Be + BO^- + B_2O_2$ | 5.85 | 5.31 | | | |
| $Mg(BO)_3^-$ | 1.215 | 2.308 | $Mg + B_3O_3^-$ | 1.47 | 1.34 | 6.48 | 6.38 | -0.50 |
| | | | $Mg + BO + B_2O_2^-$ | 6.01 | 5.09 | | | |
| | | | $Mg + BO^- + B_2O_2$ | 4.10 | 3.21 | | | |
| $Ca(BO)_3^-$ | 1.217 | 2.664 | $Ca + B_3O_3^-$ | 1.67 | 1.88 | 5.44 | 6.18 | -0.62 |
| | | | $Ca + BO + B_2O_2^-$ | 6.20 | 5.64 | | | |
| | | | $Ca + BO^- + B_2O_2$ | 4.30 | 3.75 | | | |
| $B(BO)_4^-$ | 1.212 | 1.659 | $B + B_4O_4^-$ | 7.02 | 6.70 | 8.05 | 7.81 | +0.15 |
| | | | $B + B_2O_2 + B_2O_2^-$ | 10.44 | 9.20 | | | |
| | | | $B + BO + B_3O_3^-$ | 10.62 | 9.94 | | | |
| | | | $B + BO^- + B_3O_3$ | 12.24 | 10.94 | | | |
| $Al(BO)_4^-$ | 1.209 | 2.136 | $Al + B_4O_4^-$ | 3.40 | 2.70 | 7.86 | 7.64 | -0.25 |
| | | | $Al + B_2O_2 + B_2O_2^-$ | 6.83 | 5.20 | | | |
| | | | $Al + BO + B_3O_3^-$ | 7.01 | 5.94 | | | |
| | | | $Al + BO^- + B_3O_3$ | 8.63 | 7.12 | | | |

### 3.1.1. Structures

The linear $C_{\infty v}$ structure of $M(BO)_2^-$ for M = Li, Na, and K is displayed in Fig. 3a. Table 2 lists the M-B and B-O bond lengths of $M(BO)_{k+1}^-$ anions. The M-B bond length in $M(BO)_2^-$ increases from 2.230 Å (M = Li) to 3.012 Å (M = K) as expected due to an increase in the atomic radii. However, the B-O bond length in $M(BO)_2^-$, lying in the range 1.221-1.224 Å, does not change significantly. It can be noted that the bond length of BO moiety is slightly smaller than that of BO⁻ but larger than that of BO (1.200 Å). This may suggest that the negative charge tends to be delocalized over BO moieties. Overall, the structure of M(BO)$_2^-$ closely resembles those of M(CN)$_2^-$ reported earlier for M = Li and Na [8].

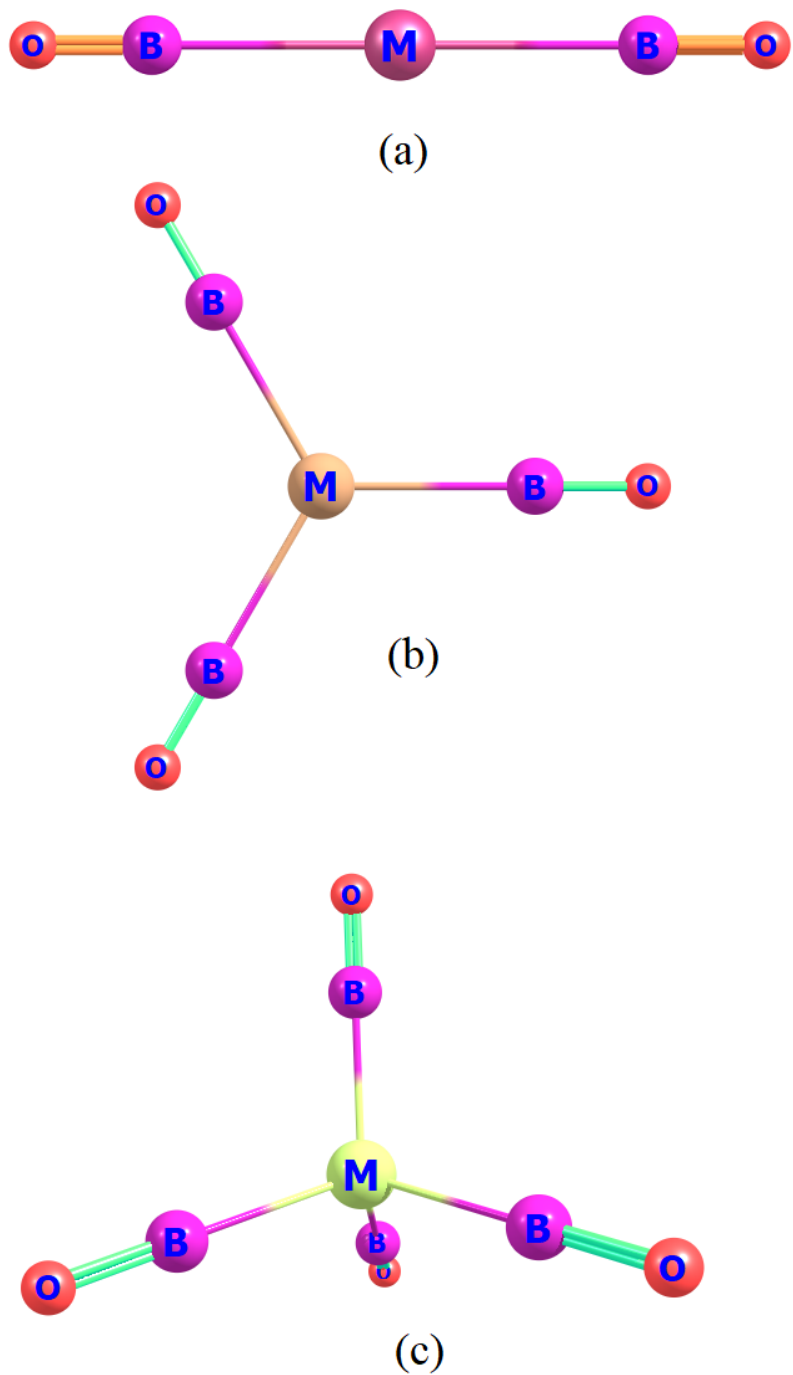

(a)

(b)

(c)

Fig. 3. B3LYP/6-311+G(3df) optimized lowest energy structures of M(BO)$_{k+1}^-$ anions for M = Li, Na, K, and $k$ = 1 (a); M = Be, Mg, Ca, and $k$ = 2 (b); M = B, Al, and $k$ = 3 (c). The corresponding bond lengths are listed in Table 2.

Fig. 3b displays the trigonal planar structure of $M(BO)_3^-$ for M = Be, Mg, and Ca, with $D_{3h}$ symmetry. In general, the bond lengths in $M(BO)_3^-$ are smaller than those of $M(BO)_2^-$ (see Table 2). Like $M(BO)_2^-$, however, the M-B bond length in $M(BO)_3^-$ increases, from 1.885 Å (M = Be) to 2.664 Å (M = Ca), with the increase in the size of M. Further, the B-O bond lengths, being in the range 1.215-1.217 Å, lie between those of BO and $BO^-$. Nevertheless, the structure of $M(BO)_3^-$ is quite similar to those of $M(CN)_3^-$ as reported previously for M = Be, Mg, and Ca [8].

The tetrahedron structure of $M(BO)_4^-$ anion for M = B and Al with $T_d$ symmetry is shown in Fig. 3c. Note that Yao et al. [28] have already reported the $T_d$ structure of $B(BO)_4^-$ in the ground-state, analogous to $BH_4^-$. Unlike $B(BO)_4^-$, the ground-state geometry of $Al(AlO)_4^-$ has been reported to be a square planar $D_{4h}$ structure [39]. The B-B and B-O bond lengths, 1.659 Å and 1.212 Å in $B(BO)_4^-$, agree with those reported by Yao et al. [28]. Likewise, the $Al(BO)_4^-$ has Al-B and B-O bond lengths of 2.123 Å and 1.209 Å, respectively (see Table 2). The structures of $M(BO)_4^-$, like $M(BO)_2^-$ and $M(BO)_3^-$, are found to be analogous to $M(CN)_4^-$ [8]. It is also noted that the B-O bond length in $M(BO)_{k+1}^-$ decreases with the increase in the valence ($k$) of M.

### 3.1.2. Stability

The stability of $M(BO)_{k+1}^-$ anions has been analyzed by considering their dissociation into M and various $(BO)_n$ fragments. The corresponding dissociation energies ($\Delta E$) are also listed in Table 2. In general, all these species are energetically stable due to positive $\Delta E$ values. However, $\Delta E$ values of $M(BO)_{k+1}^-$ decreases with the increase in the size of M for a given $k$, except in $Ca(BO)_3^-$ which possess higher $\Delta E$ values as compared to $Mg(BO)_3^-$. The $\Delta E$ values of $M(BO)_2^-$ suggest that it prefers dissociation into $B_2O_2^-$ than $BO^-$. However, $M(BO)_3^-$ favors dissociation into $B_3O_3^-$ and dissociation to $B_2O_2^-$ is less favored energetically. The $\Delta E$ values also indicate that $B(BO)_4^-$ is the most stable species among all M(BO)$_{k+1}^-$ anions considered in this study. The enhanced stability of B(BO)$_4^-$ can be expected due to the presence of sp$^3$ hybridized B center, similar to C in CH$_4$. This is why several boron clusters are reported to be highly stable [40-43]. This feature is also reflected in the highest occupied molecular orbitals (HOMOs) plotted in Fig. 4. One can see that the HOMOs of Li(BO)$_2^-$ and Be(BO)$_3^-$ are mainly contributed by BO moieties, with little or no contribution from Li and Be atoms. On the contrary, the central B atom contributes significantly to the HOMO of B(BO)$_4^-$.

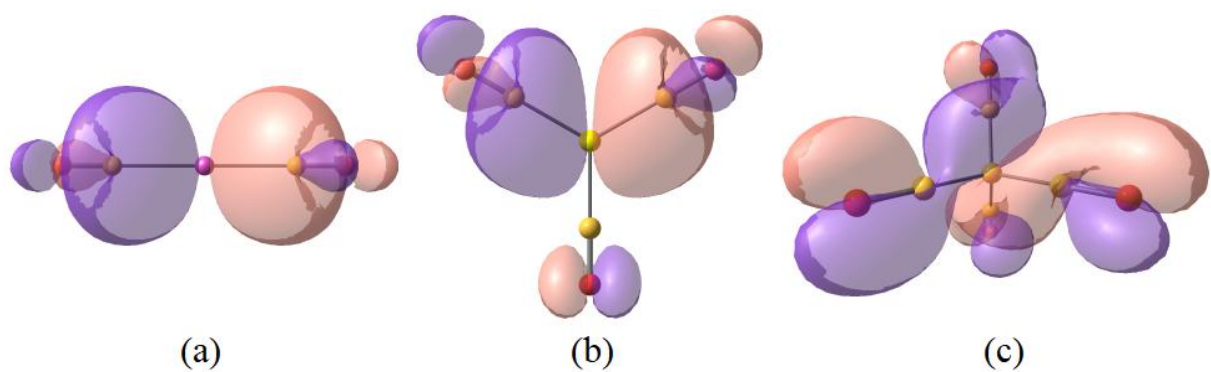

(a)                    (b)                    (c)

Fig. 4. The highest occupied molecular orbital (HOMO) of Li(BO)$_2^-$, Be(BO)$_3^-$ and B(BO)$_4^-$.

Considering the fact that thermal and entropy effects may further destabilize M(BO)$_{k+1}^-$ anions, we have also calculated and listed the Gibbs' free energy change for dissociation at 298.15 K ($\Delta G^{298}$) in Table 2. One can note that the $\Delta G^{298}$ is positive for all dissociation paths. Therefore, M(BO)$_{k+1}^-$ anions are thermodynamically stable at room temperature. Further, $\Delta G^{298}$ values of these species are large enough to suggest that M(BO)$_{k+1}^-$ can be synthesized at least via gas-phase reaction, preferably, by using BO and BO$^-$.

### 3.1.3. Superhalogen property

The superhalogen property of M(BO)$_{k+1}^-$ anions is revealed by calculating their VDEs as listed in Table 2. The VDE of M(BO)$_2^-$ ranges from 5.54 eV (M = Li) to 3.81 eV (M = K) whereas the VDE of M(BO)$_3^-$ varies from 6.73 eV (M = Be) to 5.44 eV (M = Ca). The most stable B(BO)$_4^-$ anion possesses the highest VDE of 8.05 eV, which is consistent with the reported value of 7.99 eV and 8.03 eV by the OVGF method [28]. Our calculated VDE of $Al(BO)_4^-$, 7.86 eV, is slightly smaller than those of $Al(BO_2)_4^-$ reported in the range 7.25-8.36 eV at various levels [44]. It is quite interesting to note that the VDE of $M(BO)_{k+1}^-$ anions decreases with the increase in the size of M for a fixed $k$. This fact is in contrast to previously reported superhalogens anions such as $MF_{k+1}^-$ [1], $M(CN)_{k+1}^-$ [8] but in accordance with $M(OCN)_{k+1}^-$ [8] and $M(OF)_{k+1}^-$ [12]. This fact can be explained on the basis of the relative stability of $M(BO)_{k+1}^-$ anions as discussed below. Nevertheless, the VDE of $M(BO)_{k+1}^-$ anions, being in the range 3.81-8.05 eV, are high enough to predict that these all belong to the class of superhalogens.

The superhalogen property of $M(BO)_{k+1}^-$ anions can be explained based on the charge (de)localization on BO moieties, $q(BO)$. For a given $k$, the negative charge on BO increases with the increase in the size of M. This can be expected due to the decrease in the ionization energy of M. However, the negative charge on BO decreases with the increase in the valence of M. This can be explained on the basis of charge delocalization with the increase in the BO moieties. For instance, the magnitude of $q(BO)$ in $Li(BO)_2^-$ is 0.65$e$, which is reduced to 0.27$e$. The slightly positive $q(BO)$ in $B(BO)_4^-$ is due to charge transfer to B center such that it is negatively charged (-1.60$e$) as noticed earlier [28]. The negative charge on the B center has been previously noticed by us in the $BH_6^+$ cluster [45] as well. This increase in the charge delocalization results in the increase in the VDE of $M(BO)_{k+1}^-$ anions.

As mentioned earlier, the higher VDE of $M(BO)_{k+1}^-$ anions leads to their enhanced stability. The frontier orbitals' energy gap ($E_{gap}$) is an important parameter to assess the chemical reactivity [46] and relative stability of $M(BO)_{k+1}^-$ anions. The species with higher $E_{gap}$ are less chemically less reactive and consequently, more stable. From Table 1, one can observe that the $E_{gap}$ values of these anions follow the same trend as those of their VDEs. This enables us to establish a statistical correlation between $E_{gap}$ and VDE of $M(BO)_{k+1}^-$ superhalogens. We obtain a linear correlation as shown in Fig. 5. This correlation follows the equation, VDE = 0.81128 ($E_{\text{gap}}$) + 1.21824, with a correlation coefficient ($R^2$) of 0.93888. This is probably the first statistical correlation between $E_{\text{gap}}$ and VDE of superhalogen anions, which might be useful in estimating the VDE of M(BO)$_{k+1}^-$ series of superhalogens with a given $E_{\text{gap}}$ value.

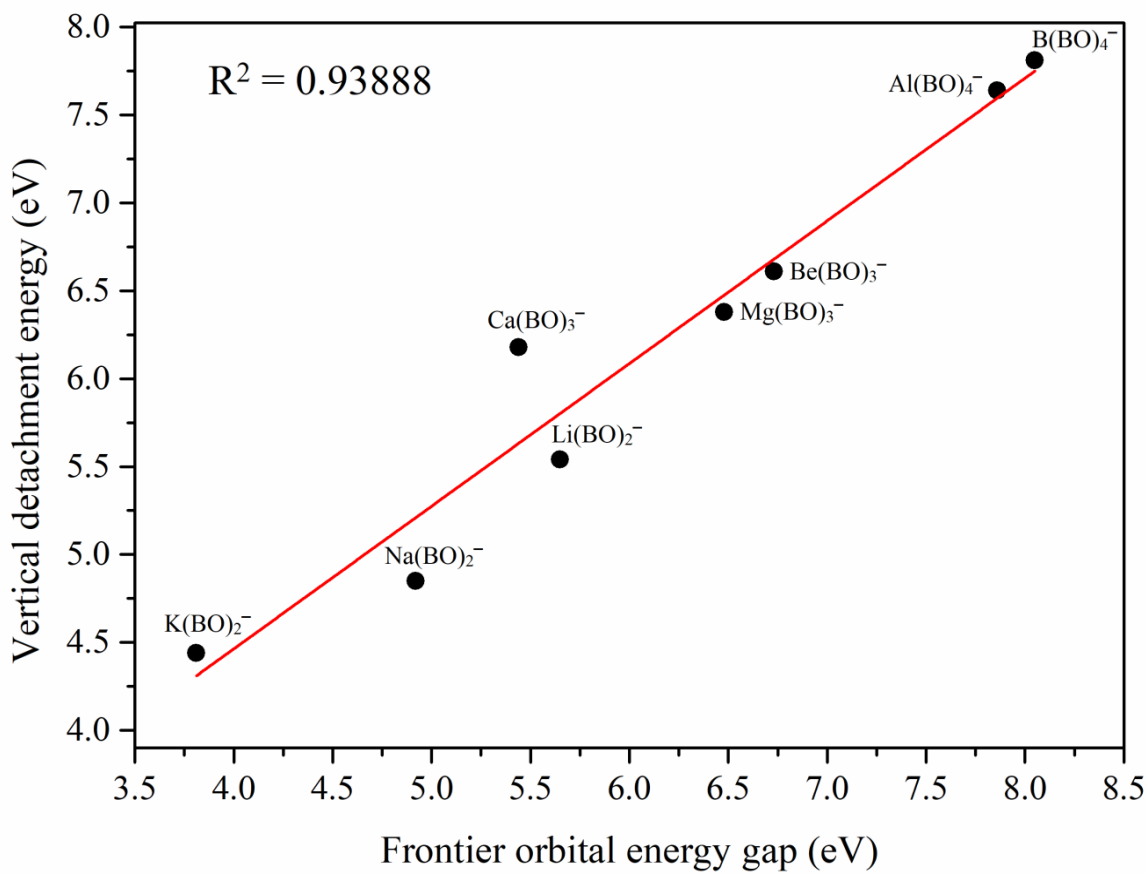

Fig. 5. The linear correlation between VDE and $E_{\text{gap}}$ of M(BO)$_{k+1}^-$ superhalogen anions.

## 3.2. BO *versus* CN ligands for superhalogens

The above discussion has established that BO can be used as a ligand to design a novel series of superhalogen anions. It has been also found that M(BO)$_{k+1}^-$ anions closely mimic the structure of M(CN)$_{k+1}^-$ species. Now, we have a look at the similarities between BO and CN ligands. Note that both are isoelectronic, which need an extra electron for shell-closure. Fig. 6 displays the structures of BO, CN, and related species. As highlighted earlier by Li et al [32], both CN and BO tend to form a dimer, just like F and the structure of $(BO)_2$ dimer closely resembles the structure of $(CN)_2$ (see Fig. 6). The $CN^-$ forms $OCN^-$ by interacting with the O atom. $OCN^-$, having a VDE of 3.64 eV, can also act as a ligand to design new superhalogens [8]. Likewise, $BO^-$ forms $BO_2^-$ superhalogen with the VDE of 4.43 eV, which can be further employed as ligands to design other complex superhalogens, also called 'hyperhalogens' [44, 47]. From Fig. 6, one can see that both $OCN^-$ and $BO_2^-$ possess linear structures. Therefore, it seems interesting to compare the potency of BO and CN as a ligand for superhalogens.

Table 3. CCSD(T)//B3LYP/6-311+G(3df) calculated vertical detachment energy (VDE) of $LiX^-$ and $X^-$ for X = BO, F, and CN. The binding energy (BE) of $X_2$ is also listed.

| $LiX^-$ anions | VDE (eV) | VDE of $X^-$ (eV) | BE of $X_2$ per X (eV) |
|---|---|---|---|
| $Li(BO)_2^-$ | 5.54 | 2.44 | 2.53 |
| $LiF_2^-$ | 6.47 | 3.22 | 0.79 |
| $Li(CN)_2^-$ | 6.85 | 3.83 | 4.05 |

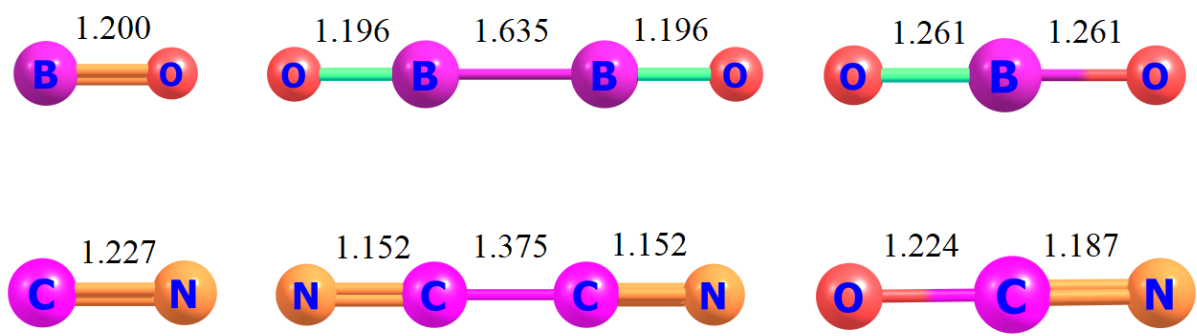

Fig. 6. The optimized structures of BO, $(BO)_2$, $BO_2^-$, CN, $(CN)_2$, and $OCN^-$ at B3LYP/6-311+G(3df) level.

To assess the potency of ligands for superhalogens, we have considered two criteria. First, whether the ligand can push up the VDE of anion beyond halogen, and the next, whether it can bind dissociatively to the central core in order to stabilize the anion. We have compared the VDE of $Li(BO)_2^-$ with those of $Li(CN)_2^-$ and $LiF_2^-$ in Table 3. One can see the VDE of $Li(BO)_2^-$ is slightly smaller than that of $LiF_2^-$ whereas $Li(CN)_2^-$ possesses a larger VDE value. This can be expected due to lower and higher VDE of $BO^-$ and $CN^-$, respectively as compared to $F^-$ (see Table 3). Based on the VDE of anion, CN seems to be a more potent ligand for superhalogen. However, it has a great tendency to form a dimer $(CN)_2$ due to higher binding energy (4.05 eV), which restricts its dissociative bonding to the central core. This feature of CN has been noticed in some previous studies [48, 49]. On the contrary, the binding energy of $(BO)_2$ dimer is calculated to be 2.53 eV (see Table 3), which is smaller than that of $(CN)_2$. Therefore, it can participate in dissociative bonding effectively. On the basis of this fact, BO can be a better ligand to stabilize superhalogen anions. The similarities between BO and CN along with its lower tendency of dimerization suggest that BO appears as an inorganic analog of CN, which can be employed to design new series of superhalogens.

## 4. Conclusions

Our CCSD(T) and B3LYP calculations suggest that BO is capable of inducing superhalogen behavior in $M(BO)_{k+1}^-$ anions as their VDE lies in the range 4.44-7.81 eV. $M(BO)_{k+1}^-$ anions are linear for M = Li, Na, K, trigonal planar M = Be, Mg, Ca, and tetrahedral for M = B, Al. These anions are energetically and thermodynamically stable against various dissociation paths. For a given $k$, their stability and VDE decrease with the increase in the size of M. This has been explained on the basis of charge delocalization on BO moieties and the highest occupied molecular orbitals. The frontier orbitals energy gap and VDE are found to be linearly correlated with the correlation coefficient of $R^2 = 0.93888$. BO ligand mimics the behavior of the CN ligand. However, its dimerization energy is lower than CN, which can make it preferable to use as an inorganic analog of CN. These findings not only add a new chapter in the research on superhalogens but also enrich the chemistry of boron.

**Acknowledgements**

The author thanks Dr. G. L. Gustev for discussion and useful suggestions. The funding received from University Grants Commission through start up grant 30-466/2019(BSR) is also acknowledged.

**Conflict of Interest**

The author declares no conflict of interests in the publication of this work.